\documentstyle[twoside,graphicx]{article}

\catcode`\@=11
\long\def\@makefntext#1{
\protect\noindent \hbox to 3.2pt {\hskip-.9pt  
$^{{\eightrm\@thefnmark}}$\hfil}#1\hfill}               

\def\thefootnote{\fnsymbol{footnote}}
\def\@makefnmark{\hbox to 0pt{$^{\@thefnmark}$\hss}}    
        
\def\ps@myheadings{\let\@mkboth\@gobbletwo
\def\@oddhead{\hbox{}
\rightmark\hfil\eightrm\thepage}   
\def\@oddfoot{}\def\@evenhead{\eightrm\thepage\hfil
\leftmark\hbox{}}\def\@evenfoot{}
\def\sectionmark##1{}\def\subsectionmark##1{}}

\oddsidemargin=\evensidemargin
\renewcommand{\thefootnote}{\fnsymbol{footnote}}

\newcounter{sectionc}\newcounter{subsectionc}\newcounter{subsubsectionc}
\renewcommand{\section}[1] {\vspace{12pt}\addtocounter{sectionc}{1} 
\setcounter{subsectionc}{0}\setcounter{subsubsectionc}{0}\noindent 
        {\tenbf\thesectionc. #1}\par\vspace{5pt}}
\renewcommand{\subsection}[1] {\vspace{12pt}\addtocounter{subsectionc}{1} 
        \setcounter{subsubsectionc}{0}\noindent 
        {\bf\thesectionc.\thesubsectionc. {\kern1pt \bfit #1}}\par\vspace{5pt}}
\renewcommand{\subsubsection}[1] {\vspace{12pt}\addtocounter{subsubsectionc}{1}
        \noindent{\tenrm\thesectionc.\thesubsectionc.\thesubsubsectionc.
        {\kern1pt \tenit #1}}\par\vspace{5pt}}
\newcommand{\nonumsection}[1] {\vspace{12pt}\noindent{\tenbf #1}
        \par\vspace{5pt}}

\newcounter{appendixc}
\newcounter{subappendixc}[appendixc]
\newcounter{subsubappendixc}[subappendixc]
\renewcommand{\thesubappendixc}{\Alph{appendixc}.\arabic{subappendixc}}
\renewcommand{\thesubsubappendixc}
        {\Alph{appendixc}.\arabic{subappendixc}.\arabic{subsubappendixc}}

\renewcommand{\appendix}[1] {\vspace{12pt}
        \refstepcounter{appendixc}
        \setcounter{figure}{0}
        \setcounter{table}{0}
        \setcounter{lemma}{0}
        \setcounter{theorem}{0}
        \setcounter{corollary}{0}
        \setcounter{definition}{0}
        \setcounter{equation}{0}
        \renewcommand{\thefigure}{\Alph{appendixc}.\arabic{figure}}
        \renewcommand{\thetable}{\Alph{appendixc}.\arabic{table}}
        \renewcommand{\theappendixc}{\Alph{appendixc}}
        \renewcommand{\thelemma}{\Alph{appendixc}.\arabic{lemma}}
        \renewcommand{\thetheorem}{\Alph{appendixc}.\arabic{theorem}}
        \renewcommand{\thedefinition}{\Alph{appendixc}.\arabic{definition}}
        \renewcommand{\thecorollary}{\Alph{appendixc}.\arabic{corollary}}
        \renewcommand{\theequation}{\Alph{appendixc}.\arabic{equation}}
        \noindent{\tenbf Appendix \theappendixc #1}\par\vspace{5pt}}
\newcommand{\subappendix}[1] {\vspace{12pt}
        \refstepcounter{subappendixc}
        \noindent{\bf Appendix \thesubappendixc. {\kern1pt \bfit #1}}
        \par\vspace{5pt}}
\newcommand{\subsubappendix}[1] {\vspace{12pt}
        \refstepcounter{subsubappendixc}
        \noindent{\rm Appendix \thesubsubappendixc. {\kern1pt \tenit #1}}
        \par\vspace{5pt}}

\newcommand{\textlineskip}{\baselineskip=13pt}
\newcommand{\smalllineskip}{\baselineskip=10pt}

\def\eightcirc{
\begin{picture}(0,0)
\put(4.4,1.8){\circle{6.5}}
\end{picture}}
\def\eightcopyright{\eightcirc\kern2.7pt\hbox{\eightrm c}} 

\newcommand{\copyrightheading}[1]
        {\vspace*{-2.5cm}\smalllineskip{\flushleft
        {\footnotesize International Journal of Modern Physics A, #1}\\
        {\footnotesize $\eightcopyright$\, World Scientific Publishing
         Company}\\
         }}

\def\abstracts#1#2#3{{
        \centering{\begin{minipage}{4.5in}\baselineskip=10pt\footnotesize
        \parindent=0pt #1\par 
        \parindent=15pt #2\par
        \parindent=15pt #3
        \end{minipage}}\par}} 

\renewenvironment{thebibliography}[1]
        {\frenchspacing
         \ninerm\baselineskip=11pt
         \begin{list}{\arabic{enumi}.}
        {\usecounter{enumi}\setlength{\parsep}{0pt}
         \setlength{\leftmargin 12.7pt}{\rightmargin 0pt} 
         \setlength{\itemsep}{0pt} \settowidth
        {\labelwidth}{#1.}\sloppy}}{\end{list}}

\newcounter{itemlistc}
\newcounter{romanlistc}
\newcounter{alphlistc}
\newcounter{arabiclistc}

\newcommand{\fcaption}[1]{
        \refstepcounter{figure}
        \setbox\@tempboxa = \hbox{\footnotesize Fig.~\thefigure. #1}
        \ifdim \wd\@tempboxa > 5in
           {\begin{center}
        \parbox{5in}{\footnotesize\smalllineskip Fig.~\thefigure. #1}
            \end{center}}
        \else
             {\begin{center}
             {\footnotesize Fig.~\thefigure. #1}
              \end{center}}
        \fi}

\newcommand{\tcaption}[1]{
        \refstepcounter{table}
        \setbox\@tempboxa = \hbox{\footnotesize Table~\thetable. #1}
        \ifdim \wd\@tempboxa > 5in
           {\begin{center}
        \parbox{5in}{\footnotesize\smalllineskip Table~\thetable. #1}
            \end{center}}
        \else
             {\begin{center}
             {\footnotesize Table~\thetable. #1}
              \end{center}}
        \fi}

\def\@citex[#1]#2{\if@filesw\immediate\write\@auxout
        {\string\citation{#2}}\fi
\def\@citea{}\@cite{\@for\@citeb:=#2\do
        {\@citea\def\@citea{,}\@ifundefined
        {b@\@citeb}{{\bf ?}\@warning
        {Citation `\@citeb' on page \thepage \space undefined}}
        {\csname b@\@citeb\endcsname}}}{#1}}

\newif\if@cghi
\def\cite{\@cghitrue\@ifnextchar [{\@tempswatrue
        \@citex}{\@tempswafalse\@citex[]}}
\def\citelow{\@cghifalse\@ifnextchar [{\@tempswatrue
        \@citex}{\@tempswafalse\@citex[]}}
\def\@cite#1#2{{$\null^{#1}$\if@tempswa\typeout
        {IJCGA warning: optional citation argument 
        ignored: `#2'} \fi}}

\def\pmb#1{\setbox0=\hbox{#1}
        \kern-.025em\copy0\kern-\wd0
        \kern.05em\copy0\kern-\wd0
        \kern-.025em\raise.0433em\box0}

\def\fnt#1#2{\footnotetext{\kern-.3em
        {$^{\mbox{\scriptsize #1}}$}{#2}}}

\def\fpage#1{\begingroup
\voffset=.3in
\thispagestyle{empty}\begin{table}[b]\centerline{\footnotesize #1}
        \end{table}\endgroup}

\font\tenrm=cmr10
\font\tenit=cmti10 
\font\tenbf=cmbx10
\font\bfit=cmbxti10 at 10pt
\font\ninerm=cmr9

\font\eightrm=cmr8

\newcommand{\Ab}        {\mbox{$\mathrm{A}$}}
\newcommand{\hb}        {\mbox{$\mathrm{h}$}}
\newcommand{\Hb}        {\mbox{$\mathrm{H}$}}

\newcommand{\mha}       {\mbox{($m_{\mathrm{h}}$,$m_{\mathrm{A}}$)}}
\newcommand{\mhtgb}       {\mbox{($m_{\mathrm{h}}$,$ \tan\beta $)}}
\newcommand{\matgb}       {\mbox{($m_{\mathrm{A}}$,$ \tan\beta $)}}
\newcommand{\mtop} {\mbox{$m_{\mathrm{t}}$}}
\newcommand{\ra}    {\rightarrow}
\newcommand{\tanb}  {\mbox{$ \tan\beta $}} 
\newcommand{\zboson}    {\mbox{$\mathrm{ Z }$}}
\newcommand{\ee}      {\mbox{$\mathrm{e^+ e^-}$}}
\newcommand{\mh}        {\mbox{$m_{\mathrm{h}}$}}
\newcommand{\mA}        {\mbox{$m_{\mathrm{A}}$}}

\def\qed{\hbox{${\vcenter{\vbox{                        
   \hrule height 0.4pt\hbox{\vrule width 0.4pt height 6pt
   \kern5pt\vrule width 0.4pt}\hrule height 0.4pt}}}$}}

\renewcommand{\thefootnote}{\fnsymbol{footnote}}        

\begin{document}
\begin{titlepage}
\def\thefootnote{\fnsymbol{footnote}}       

\begin{center}
\mbox{ } 

\vspace*{-4cm}


\end{center}
\begin{flushright}
\Large
\mbox{\hspace{10.2cm} hep-ph/0011285} \\
\mbox{\hspace{10.2cm} IEKP-KA/2000-22} \\
\mbox{\hspace{10.2cm} November 2000}
\end{flushright}
\begin{center}
\vskip 2.0cm
{\Huge\bf
A General MSSM
\vskip 1.0cm
Parameter Scan}
\vskip 2.5cm
{\LARGE\bf Andr\'e Sopczak}\\
\smallskip
\vskip 1cm
\large University of Karlsruhe

\vskip 2.5cm
\centerline{\Large \bf Abstract}
\end{center}

\vskip 2.5cm
\hspace*{-3cm}
\begin{picture}(0.001,0.001)(0,0)
\put(,0){
\begin{minipage}{16cm}
\Large
\renewcommand{\baselinestretch} {1.2}
The excluded $\tan\beta$ range and Higgs boson mass
regions in the framework of the Minimal Supersymmetric 
extension of the Standard Model (MSSM) depend on several 
parameters. 
The Higgs boson masses, cross-sections and branching 
fractions have been determined including two-loop 
diagrammatic calculations.
The limits obtained with a more general scan over the parameter 
space of the MSSM are compared with those in the so-called 
benchmark scenario.
The combination of the searches for Higgs particles in the 1999
data collected by the DELPHI collaboration at center-of-mass energies 
between 191.6 and 201.7 GeV allows stringent limits to be set in 
combination with previous DELPHI results.
In addition, an interpretation in the framework of the general MSSM scan
of the 2000 LEP data at the hightest energies between 201.7 and 209.0 GeV  
is given. We show that the current data for the HZ and hA production 
can be comfortably accommodated in the MSSM.
\renewcommand{\baselinestretch} {1.}

\normalsize
\vspace{1.5cm}
\begin{center}
{\sl \large
Presented at the DPF--2000, Columbus, Ohio, USA, Aug. 9--12, 2000
\vspace{-6cm}
}
\end{center}
\end{minipage}
}
\end{picture}
\vfill

\end{titlepage}


\newpage
\thispagestyle{empty}
\mbox{ }
\newpage
\setcounter{page}{1}


\normalsize\textlineskip
\thispagestyle{empty}
\setcounter{page}{1}

\copyrightheading{}                     

\vspace*{0.88truein}

\fpage{1}
\centerline{\bf A General MSSM Parameter Scan}
\vspace*{0.37truein}
\centerline{\footnotesize ANDR\'E SOPCZAK\footnote{On behalf of the DELPHI Collaboration.
            E-mail: andre.sopczak@cern.ch}}
\vspace*{0.015truein}
\centerline{\footnotesize\it Institut f\"ur experimentelle Kernphysik,
Universit\"at Karlsruhe,Wolfgang-Gaede-Strasse 1}
\baselineskip=10pt
\centerline{\footnotesize\it D-76128 Karlsruhe
Germany}

\vspace*{0.21truein}
\abstracts{
The excluded $\tan\beta$ range and Higgs boson mass
regions in the framework of the Minimal Supersymmetric 
extension of the Standard Model (MSSM) depend on several 
parameters. 
The Higgs boson masses, cross-sections and branching 
fractions have been determined including two-loop 
diagrammatic calculations.
The limits obtained with a more general scan over the parameter 
space of the MSSM are compared with those in the so-called 
benchmark scenario.
The combination of the searches for Higgs particles in the 1999
data collected by the DELPHI collaboration at center-of-mass energies 
between 191.6 and 201.7 GeV allows stringent limits to be set in 
combination with previous DELPHI results.
In addition, an interpretation in the framework of the general MSSM scan
of the 2000 LEP data at the hightest energies between 201.7 and 209.0 GeV  
is given. We show that the current data for the HZ and hA production 
can be comfortably accommodated in the MSSM.
}{}{}


\vspace*{1pt}\textlineskip      
\section{Introduction}  
\vspace*{-0.5pt}
\noindent
The importance of a general MSSM parameter scan has already been pointed
out~\cite{jras}. For earlier LEP2 data taken up to $\sqrt{s} = 172$~GeV  
it was shown that the benchmark limits on the pseudoscalar Higgs mass 
disappeared completely~\cite{as97}. 
The data taken by DELPHI in 1997 and 1998 up to $\sqrt{s} = 189$~GeV
set strong limits on the masses of the neutral Higgs bosons, but they 
were 6 to $8~{\rm GeV/c^2}$ lower than the benchmark limits~\cite{delphi98}.
Important parameters for the Higgs boson phenomenology are~\cite{jras}:

\begin{itemize}
\item $m_{\rm h}$ or $m_{\rm A}$ -- the Higgs boson mass values investigated.
\vspace*{-0.3cm}
\item \tanb\ -- the ratio between the two Higgs vacuum expectation values.
\vspace*{-0.3cm}
\item $m_{\mathrm{sq}}$ -- the common mass parameter for all squarks at the electroweak scale.
\vspace*{-0.3cm}
\item $M_2$ -- the common SU(2) gaugino mass parameter at the electroweak scale.
\vspace*{-0.3cm}
\item $\mu$ -- the mixing parameter of the Higgs doublets in the superpotential.
\vspace*{-0.3cm}
\item $A$ -- the stop mixing term.
The mixing term is defined as $X_t = Am_{\mathrm{sq}} - \mu / \tan\beta$.
\end{itemize}

For this analysis, the parameters shown in 
Table~\ref{tab:allparameters} are the input parameters 
for the calculations of the physical Higgs, sfermion, chargino, and
neutralino masses. 
These parameters were varied in the ranges shown in
Table~\ref{tab:allparameters}.
For each $m_{\mathrm{A}}$, 2700 parameter combinations were investigated.
Compared to previous studies~\cite{delphi98}, 
the large $\mu$ scenario ($\rm \mu=\pm 1000~GeV/c^2$) and a very small $M_2$ value 
($M_2=70~{\rm GeV/c^2}$) were investigated with the result that the mass limit is not 
affected.
For large $\mu$ values, a few parameter combinations were found where
the branching $\rm h\ra b\bar{b}$ vanishes;
those points were excluded by flavour independent searches.

\begin{table}[htb]
\vspace*{-0.5cm}
\begin{center}
\begin{tabular}{|c|c|c|c|c|c|}
\hline
$m_{\mathrm{A}}$ (${\rm GeV/c^2}$) & \tanb &
$m_{\mathrm{sq}}$ (${\rm GeV/c^2}$)
& $M_{\mathrm{2}}$  (${\rm GeV/c^2}$) & $\mu$ (${\rm GeV/c^2}$) & $A$ \\
\hline
20\,---\,1000 & $0.5$\,---\,50 & 200\,---\,1000 &
200\,---\,1000
& $-500$\,---\,+500 & $-2$\,---\,$+2$ \\
\hline
\end{tabular}
\end{center}
\vspace*{-0.1cm}
\tcaption{\label{tab:allparameters}
Ranges of SUSY parameters at the electroweak scale used for independent
variation in the
study of the MSSM neutral Higgs boson searches.}
\vspace*{-0.5cm}
\end{table}

Throughout this study, the top quark mass is fixed at 
$\mtop= 175 \, {\rm GeV/c^2}$.
Besides the direct exclusions from Higgs boson searches,
the following constraints have been studied: 
${\mathrm b} \ra {\mathrm s}\gamma$, 
the electroweak parameter
$\Delta \rho = \alpha_{\rm em}T_{\rm MSSM}$, and
chargino and neutralino mass limits from direct searches.
These constraints have little influence on the excluded parameter
regions with the present Higgs mass limits~\cite{delphi98}. 
Therefore, the excluded parameters have been determined from the Higgs boson 
searches alone.

\section{Excluded \mha, \mhtgb\ and \matgb\ Regions}

The results are presented in the \mha\ plane
as well as in the planes \mhtgb\ and \matgb.
In the mass plane \mha\, the A mass was scanned in steps of 1~${\rm GeV/c^2}$
up to 200~${\rm GeV/c^2}$.
In addition, A masses between 200 and 1000~${\rm GeV/c^2}$ were explored
with larger step sizes.
For each mass combination, the cross-sections of the reactions
$\ee\ra\hb\zboson,\Hb\zboson$, $\ee\ra\hb\Ab,\Hb\Ab$,
and the branching ratios for \hb\ and \Ab\ decays were
computed as functions of the parameters described in 
Table~\ref{tab:allparameters}.
Both \hb\ production through the bremsstrahlung process and
$\hb\Ab$ pair-production processes were taken into 
account~\cite{DELPHIh99}.

For some parameter combinations the branching ratio into a pair
of neutralinos is dominant. In such a case no limit can be derived
using the above listed search channels. Therefore, limits from the
DELPHI search for invisible decays of neutral Higgs bosons~\cite{DELPHIh2000}
are applied.
In addition, the decay $\rm h\ra AA$ is allowed for a small set
of parameters and taken into account for setting the limits.

A given \mha\ combination is excluded if for all
SUSY parameter sets (from the ranges defined in
Table~\ref{tab:allparameters} and for fixed $\mtop=175 \, {\rm GeV/c^2}$)
the exclusion confidence level is larger than 95\% in the combination
of all search channels~\cite{alex}.

\begin{figure}[htbp]
\vspace*{-0.3cm}
\begin{minipage}[t]{0.5\linewidth}
\centering
\includegraphics[scale=0.3]{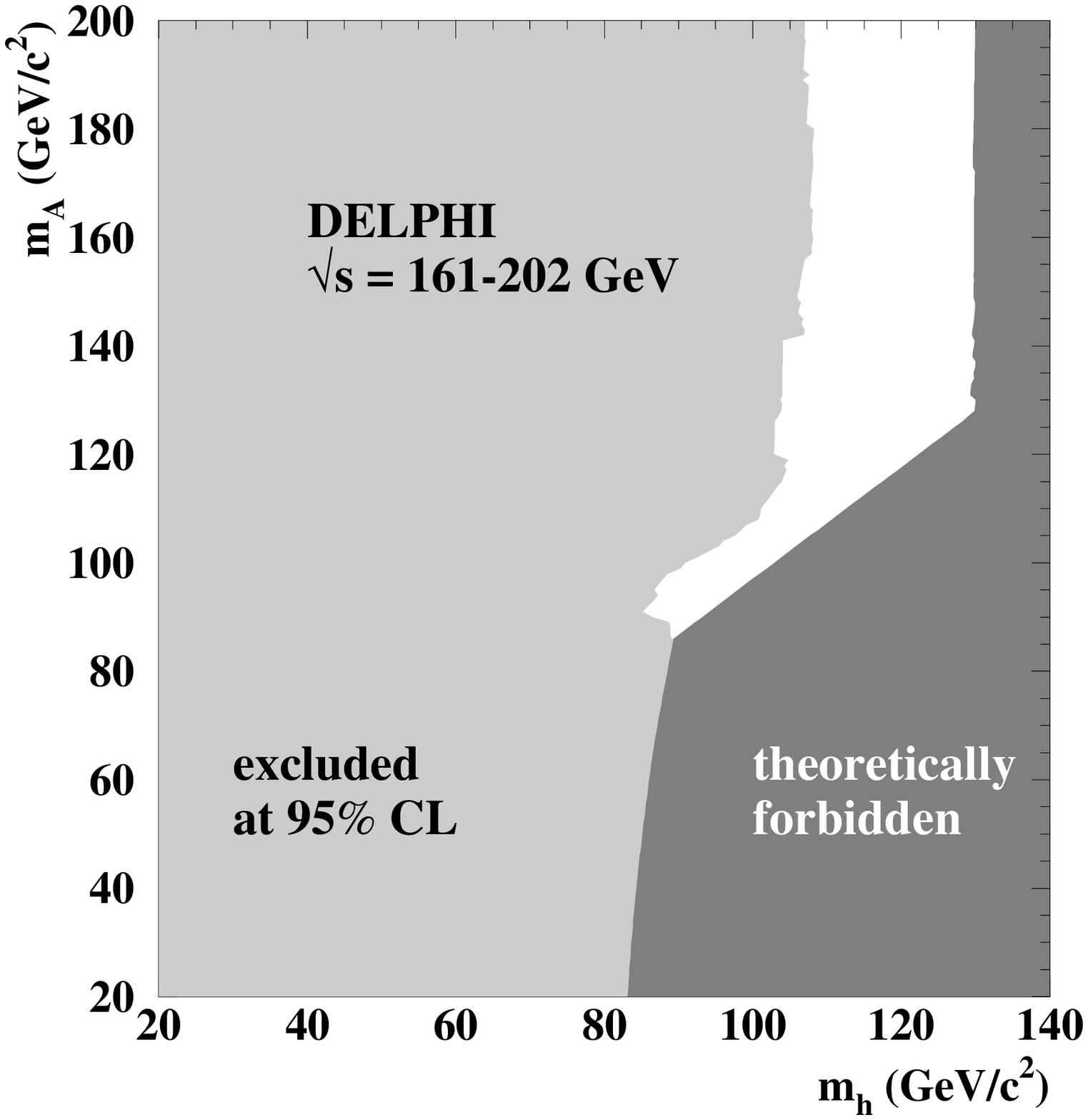}
\end{minipage}%
\begin{minipage}[t]{0.5\linewidth}
\centering
\includegraphics[scale=0.3]{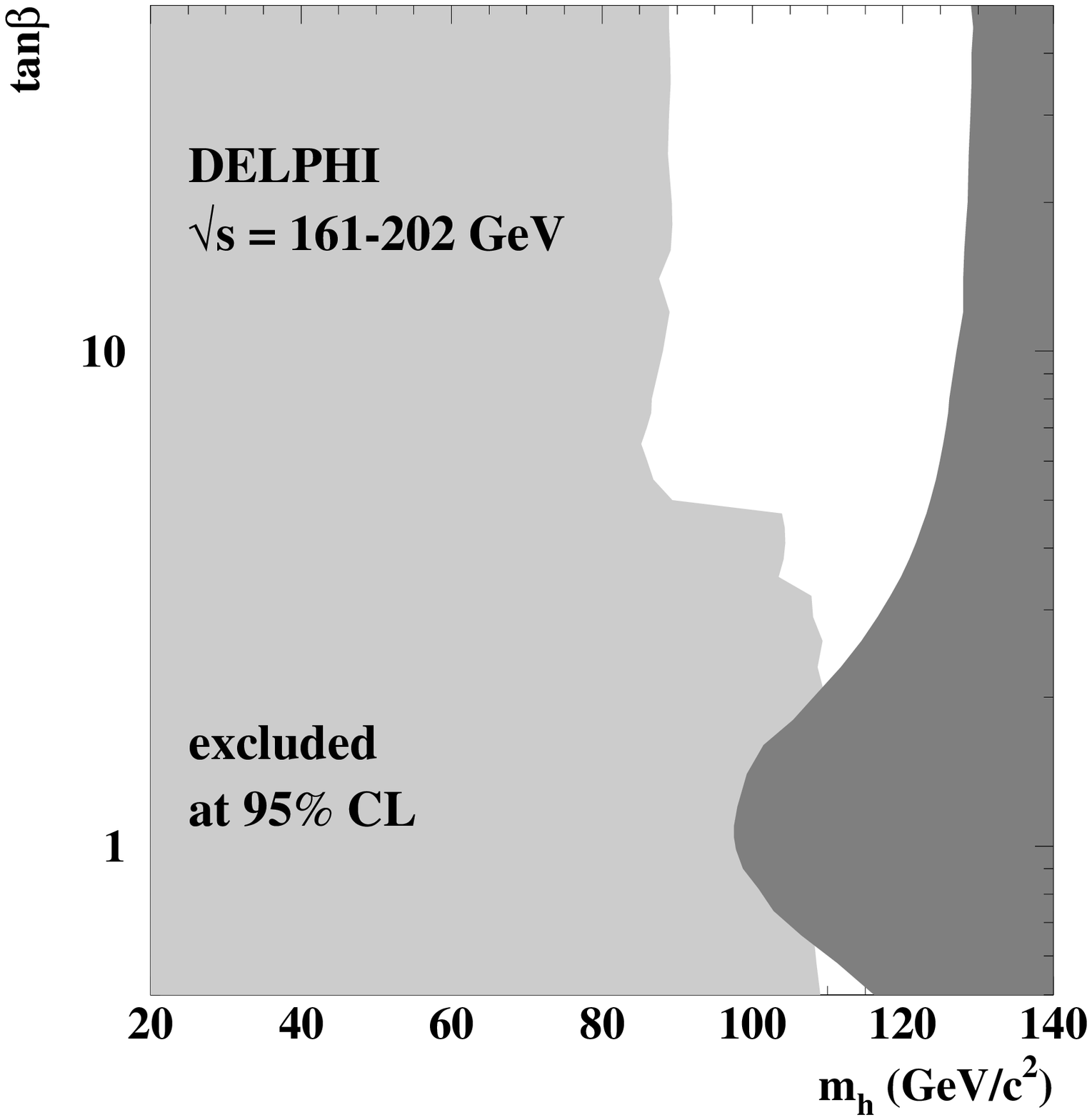}   
\end{minipage}
\vspace*{-0.5cm}
%
\begin{minipage}[t]{0.5\linewidth}
\centering
\includegraphics[scale=0.3]{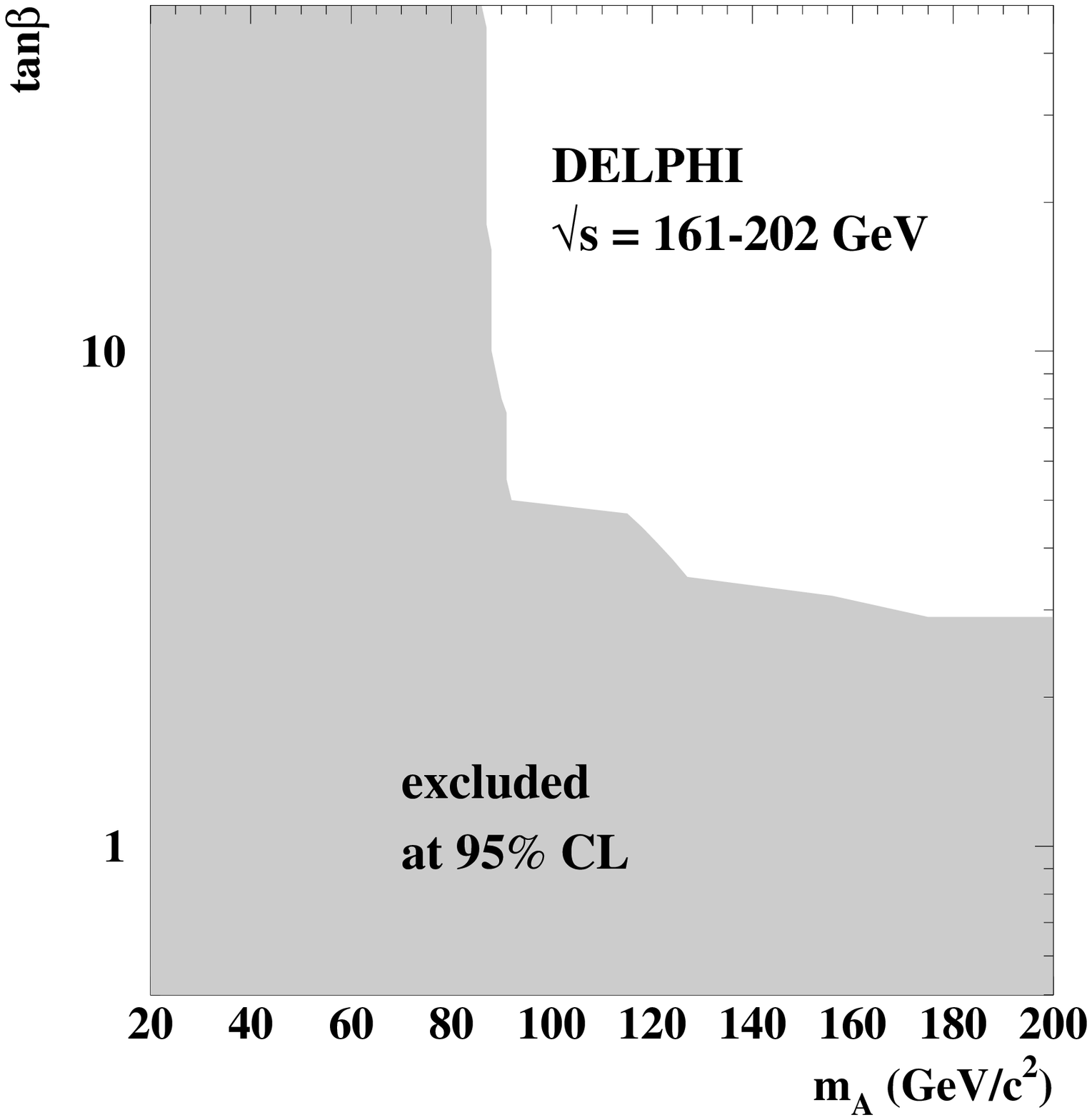}
\fcaption{\label{fig1} $\mha$, $\mhtgb$ and $\matgb$ exclusions for preliminary 
1999 DELPHI data.}
\end{minipage}%
\begin{minipage}[t]{0.5\linewidth}
\centering
\vspace*{-3.4cm}
\begin{tabular} {|c|c|c|c|c|}\hline
$m_{\rm A}$ & $m_{\rm h}$ &$m_{\rm H}$ & $\tan\beta$ & $m_{\rm sq}$ \\ 
90  & 90.0  & 114.0 & 16 & 1000 \\
100 & 99.3  & 114.0 & 16 & 1000 \\ \hline
$M_{\rm 2}$ & $\mu$       &$A$         & $\sigma_{\rm HZ}^{208}$ & $\sigma_{\rm hA}^{208}$ \\
500 & 500  &  0 & 118 & 44 \\
500 & -500 &  0 &  97 &  8 \\ \hline
\end{tabular}
\vspace*{0.4cm}
\tcaption{\label{tab:combi}
\mbox{Examples of parameter combinations} \newline
\mbox{in the MSSM which combine excesses in the} \newline
\mbox{2000 LEP data. All masses are given in $\rm GeV/c^2$} \newline
and cross-sections in fb.}
\end{minipage}
\end{figure}

Figure~\ref{fig1} shows three regions:
the 95\% CL excluded region (light grey), the theoretically not allowed 
region (dark), and the allowed region (white).

The extended parameter range results in a smaller excluded mass region 
compared to those obtained with the benchmark~\cite{DELPHIh99}.
However, because of the large statistics, the mass limits are only 1 to 
2~GeV lower.
\vspace*{-1mm}

\section{Combining $m_{\rm H}=114$~GeV and $\mh\approx\mA\approx 90$~GeV}

An excess has been observed in the 2000 LEP Higgs data
for the HZ production~\cite{lephiggs},
and we note that the mass limits from hA production are about 2 GeV
below the expectation, however, no significant hA excess is 
claimed~\cite{osaka}.
A scan over the allowed MSSM parameter combinations shows that a 
possible interpretation can be achieved when all three
neutral Higgs bosons h, A and H are light.
Table~\ref{tab:combi} gives parameter combinations which 
could explain the HZ and the hA results and which are also consistent 
with the excluded range from 1999 LEP data based on the general 
scan described before~\cite{as99}.
The HZ production cross section at $\sqrt{s} = 208$~GeV 
$\sigma^{208}_{\rm HZ}$ is approximately the Standard Model 
value and thus the expected signal event rate would agree 
with the observed excess.
\vspace*{-6mm}

\newpage
\section{Conclusions}

With the large statistics from the 1999 DELPHI data,
a general scan over the MSSM parameter space results in the following mass
limits: $\mh>85~{\rm GeV/c^2}$ and $\mA>86~{\rm GeV/c^2}$.
The range $0.8< \tanb < 1.7$ is also excluded at 95\% CL.
We have shown that the combined 2000 LEP data for Higgs boson 
bremsstrahlung and Higgs boson pair-production fits well
in the MSSM for large $\tan\beta$ values.


\vspace*{-0.25cm}
\nonumsection{References}

\end{document}